\documentclass{SciPost}

\hypersetup{
    colorlinks,
    linkcolor={red!50!black},
    citecolor={blue!50!black},
    urlcolor={blue!80!black}
}

\usepackage[bitstream-charter]{mathdesign}
\DeclareSymbolFont{usualmathcal}{OMS}{cmsy}{m}{n}
\DeclareSymbolFontAlphabet{\mathcal}{usualmathcal}

\fancypagestyle{SPstyle}{
\fancyhf{}
\lhead{\colorbox{scipostdeepblue}{\bf \color{white} ~SciPost Physics Proceedings }}
\rhead{{\bf \color{scipostdeepblue} ~Submission }}

\fancyfoot[C]{\textbf{\thepage}}
}
\makeatletter
\let\c@lofdepth\relax
\let\c@lotdepth\relax
\makeatother
\usepackage{subfigure}

\begin{document}

\pagestyle{SPstyle}

\begin{center}{\Large \textbf{\color{scipostdeepblue}{
%%%%%%%%%% TODO: Write your article's title here
Architecturally Constrained Solutions to Ill-Conditioned Problems in QUBIC\\
%%%%%%%%%% END TODO: TITLE
}}}\end{center}

\begin{center}\textbf{
%%%%%%%%%% TODO: AUTHORS
% Write the author list here. 
% Use (full) first name (+ middle name initials) + surname format.
% Separate subsequent authors by a comma, omit comma and use "and" for the last author.
% Mark the corresponding author(s) with a superscript symbol in this order
% \star, \dagger, \ddagger, \circ, \S, \P, \parallel, ...
Leonora Kardum 1\textsuperscript{1$\star$}
%%%%%%%%%% END TODO: AUTHORS
}\end{center}

\begin{center}
%%%%%%%%%% TODO: AFFILIATIONS
% Write all affiliations here.
% Format: institute, city, country
{\bf 1} Laboratoire Astroparticule et Cosmologie (APC), Université Paris-Cité, Paris, France
\\

%%%%%%%%%% END TODO: AFFILIATIONS
%%%%%%%%%% TODO: EMAIL
% Provide email address of corresponding author(s)
$\star$ \href{mailto:email1}{\small kardum@apc.in2p3.fr}\,\quad
%%%%%%%%%% END TODO: EMAIL
\end{center}

\definecolor{palegray}{gray}{0.95}
\begin{center}
\colorbox{palegray}{
  \begin{tabular}{rr}
  \begin{minipage}{0.37\textwidth}
    \includegraphics[width=60mm]{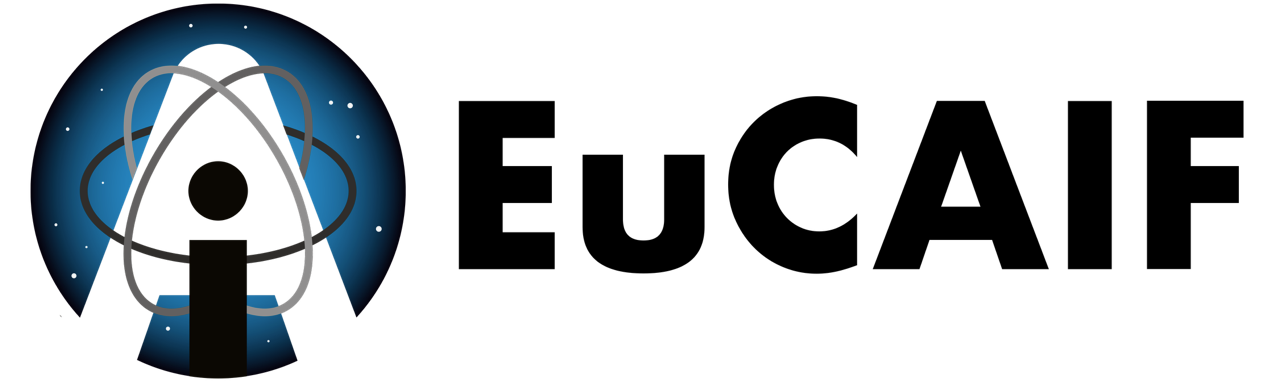}
  \end{minipage}
  &
  \begin{minipage}{0.5\textwidth}
    \vspace{5pt}
    \vspace{0.5\baselineskip} 
    \begin{center} \hspace{5pt}
    {\it The 2nd European AI for Fundamental \\Physics Conference (EuCAIFCon2025)} \\
    {\it Cagliari, Sardinia, 16-20 June 2025
    }
    \vspace{0.5\baselineskip} 
    \vspace{5pt}
    \end{center}
    
  \end{minipage}
\end{tabular}
}
\end{center}

\section*{\color{scipostdeepblue}{Abstract}}
\textbf{\boldmath{%
%%%%%%%%%% TODO: ABSTRACT
% Write your abstract here. The abstract is in boldface, and should fit in 8 lines. It should be written in a clear and accessible style, emphasizing the context, the problem(s) studied, the methods used, the results obtained, the conclusions reached, and the outlook. You can add a table contents, recommended if your paper is more than 6 pages long.
This article introduces a new physics-guided Machine Learning framework, with which we solve the generally non-invertible, ill-conditioned problems through an analytical approach and constrain the solution to the approximate inverse with the architecture of Neural Networks. By informing the networks of the underlying physical processes, the method optimizes data usage and enables interpretability of the model while simultaneously allowing estimation of detector properties and the propagation of their corresponding uncertainties. The method is applied in reconstructing Cosmic Microwave Background (CMB) maps observed with the novel interferometric QUBIC experiment aimed at measuring the tensor-to-scalar ratio r.  
%%%%%%%%%% END TODO: ABSTRACT
}}

\vspace{\baselineskip}

%%%%%%%%%% BLOCK: Copyright information
% This block will be filled during the proof stage, and finilized just before publication.
% It exists here only as a placeholder, and should not be modified by authors.
\noindent\textcolor{white!90!black}{%
\fbox{\parbox{0.975\linewidth}{%
\textcolor{white!40!black}{\begin{tabular}{lr}%
  \begin{minipage}{0.6\textwidth}%
    {\small Copyright attribution to authors. \newline
    This work is a submission to SciPost Phys. Proc. \newline
    License information to appear upon publication. \newline
    Publication information to appear upon publication.}
  \end{minipage} & \begin{minipage}{0.4\textwidth}
    {\small Received Date \newline Accepted Date \newline Published Date}%
  \end{minipage}
\end{tabular}}
}}
}
%%%%%%%%%% BLOCK: Copyright information

%%%%%%%%%% TODO: LINENO
% For convenience during refereeing we turn on line numbers:
%\linenumbers
% You should run LaTeX twice in order for the line numbers to appear.
%%%%%%%%%% END TODO: LINENO

%%%%%%%%% TODO: CONTENTS 
% Write your article contents here, starting from first \section.
% An example structure is given below.

\section{Introduction}
\label{sec:intro}
% TODO: write your article here.
\par
The detection of B-modes and constraint of the tensor-to-scalar ratio r from the measurement of Cosmic Microwave Background (CMB) would confirm the existence of primordial gravitational waves. However, recovering CMB from the data of a ground-based experiment is an inverse problem, due to the complex nature of the foregrounds and the instrumental effects.
\par
The QUBIC experiment (Q \& U Bolometric Interferometer for Cosmology) is a bolometric interferometer combining the advantages of both technologies in its efforts to detect B-modes of CMB polarization \cite{qubicI} \cite{qubicII}. The measurement process is a series of nontrivial instrumental operations, including modulation by a half-wave rotating plane and interferometric projection of the scanned sky. Most of the operators describing this process are ill-conditioned and non-invertible, which makes the solution to the inverse problem numerically (and analytically) challenging.
\par 
Conventional data-driven machine learning models offer fast and precise solutions to inversions, but usually lack interpretability and cannot be generalized between different instrumental configurations or to similar, related problems. Here we propose a novel approach to physics-guided machine learning that uses the knowledge of the physical process to build the inverse model and constraints it using neural network architectures.
\par
On top of convenient inversion of given operators, this approach allows for learning variable instrumental parameters and propagation of now explainable uncertainties to the final estimate. This framework is demonstrated here on CMB map reconstruction in the context of QUBIC. 

\section{Operator-based Modelling of the QUBIC Instrument}
\label{sec:qubic}
QUBIC  is a bolometric interferometer aimed at measuring the polarization of the CMB signal, and is located at Salta, Argentina at 4900 meters above sea level, in dry atmospheric conditions great for millimeter wave observations. The combination of interferometry and bolometry allows QUBIC to perform spectral imaging \cite{mapmaking} \cite{mapmakingII} and results in a complex acquisition and reconstruction process.
\par
\begin{figure}
    \centering
    \subfigure[]{\includegraphics[width=0.54\textwidth]{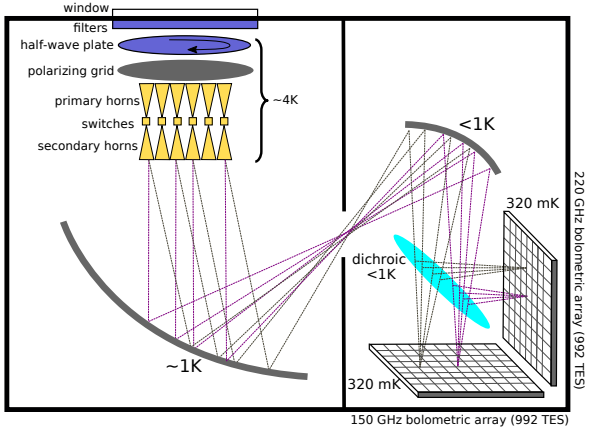}} 
    \subfigure[]{\includegraphics[width=0.32\textwidth]{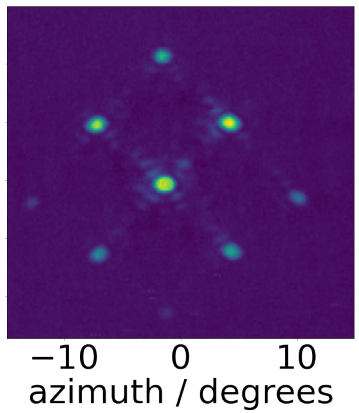}} 
    \caption{(a)  Schematic of the QUBIC instrument. (b) Synthesized beam, the interferometric pattern, at 150 GHz with 9 peaks. Taken from \cite{qubicIII}. }
    \label{fig:qubic}
\end{figure}

The conversion of the observed sky $s$ to the measured Time Ordered Data (TOD) $d$ 
\begin{equation}
d = Hs + n.
\end{equation}
is modeled with a series of operators, where each describes a specific effect in the acquisition process happening in the instrument depicted on Figure \ref{fig:qubic}. This includes the convolution with the synthesized beam, polarizer, the bolometer response, filtering, and others. The full forward model is given by the composition operator $H$ 
\begin{equation}
\label{eq:H}
H = B \cdot I \cdot D \cdot L \cdot  Hw \cdot  P \cdot  F \cdot A \cdot T \cdot U
\end{equation}
that maps multifrequency multipolarization skies to TODs for each of 992 QUBIC detectors. The instrument scans the sky with a multipeak interferometric pattern shown in Figure \ref{fig:qubic}.
\par
This is used in simulating realistic QUBIC TODs for given skies, as well as to reconstruct skies from measured TODs. The reconstruction usually relies on iterative optimization in forward modeling, where the minimization of the cost function is done by a Preconditioned Conjugate Gradient (PCG) \cite{pcg}. Apart from requiring repeated application of forward and transpose operators which can be memory expensive, each sky map reconstruction requires iterative optimization that may take hours and cannot be quickly reapplied to a new set of TODs. This motivates for a new approach to sky map reconstruction using modular neural networks.

\section{Modular Neural Architecture with Embedded Physics}
Instead of conventionally training the neural network on pairs of TODs and maps in the data-driven approach, the network is built as a series of modules where each corresponds to either the exact or the approximate inverse of known instrumental operators. The physics-guided inversions are embedded directly into the solution using the architecture of the neural network. The forward model of QUBIC acquisition is given with Equation \ref{eq:H} whose inverse does not exist, which maps any given sky to a TOD observed with the instrument. Therefore, we state there exists an operator that maps observed TODs back to given skies, which has to have the form of 
\begin{equation}
\label{eq:inverseH}
\tilde{H}^{-1} = U^{-1} \cdot T^{-1} \cdot A^{-1} \cdot F^{-1} \cdot  P^{-1} \cdot  Hw^{-1} \cdot  L^{-1} \cdot D^{-1}  \cdot I^{-1} \cdot B^{-1}. 
\end{equation}
\par
Operators from Equation \ref{eq:inverseH} with a known analytical inverse are implemented as \textit{deterministic} or \textit{dynamic} layers, while the ill-conditioned inverses are differently treated in modules which we call \textit{learnable}. 

\subsection{Examples of operator inversion}
\label{subsec:examples}

\paragraph{Bolometer time response as a dynamic module.}  
The bolometer decay convolves the signal with a truncated exponential with some decay constant $\tau$
\begin{equation}
\label{eq:bolometer}
d(t) = \int d(t') e^{-(t-t')/\tau}\, dt'.
\end{equation}
and is a noninvertible operator in time-space. However, it becomes invertible after transformation to the Fourier space with the solution
\begin{equation}
\label{eq:inversebolometer}
B^{-1}(\omega) = (1+i\omega\tau).
\end{equation}
This inverse is then implemented into the architecture as a linear layer with fixed kernel weights
\begin{equation}
\label{eq:inversebolometerlayer}
\mathbf{B^{-1}} = \texttt{nn.Linear} \left( 1 + i\omega \cdot \texttt{nn.Parameter}(\Theta_\tau) \right)
\end{equation}
and an adjustable parameter which is fitted during backpropagation. The layer is differentiable and becomes physically interpretable since the adjusted parameter corresponds to the decay constant. The main advantage is that the network is constrained by the layer architecture and cannot divert to different, non-physical solutions other than the analytical one.

%\paragraph{Polarizer as a deterministic module. (i would actually kick this out because the dynamic example is enough in my opinion and move to projection, or add something more fun like transmission)}  
%The QUBIC signal is modulated by the half-wave plate \cite{qubicVI} and filtered by the polarizer grid. The two operators act on skies as
%\begin{equation}
%\label{eq:polarizer}
%L(t) Hw(t) \cdot 
%\begin{bmatrix}
%I \\ Q \\ U
%\end{bmatrix}
%=
%0.5\, [1\;\; \cos 4\psi(t)\;\; \sin 4\psi(t)] \cdot
%\begin{bmatrix}
%I \\ Q \\ U
%\end{bmatrix}
%\end{equation}
%where $\psi(t)$ is the half-wave plate angle at some point $t$, making each TOD sample a linear combination of Stokes parameters. This operator is rank-deficient and noninvertible. A single TOD measurement cannot be used to recover three unknowns, but combining four measurements at the same pointing with different $\psi$ angles produces a $4x3$ system which is full-rank, therefore allowing a deterministic inversion and reconstruction of all Stokes at some pointed (pointed at? observed? scanned?) location on the sky. 

\paragraph{Projection operator as a learnable module.}
Each TOD sample is a sum of nine different peaks with weights $w_m$ of the interferometric synthesized beam shown in Figure \ref{fig:qubic} where different sky pixels $s_n$ are mixed into the same sample by
\begin{equation}
\label{eq:todsum}
d_j = \sum_{k = 9} w_{jk} \, s_k.
\end{equation}
Each projection is represented by a graph \cite{graphs} which keeps the relational information among all the observed pixels. It is important to note that these pixels are usually not neighbours on the sphere but are rather distant parts of the sky that are \textit{coobserved} at the same time with the interferometric pattern. Due to the collapse of multiple to one, as in Equation \ref{eq:todsum}, the projection is noninvertible and the $P^{T}P$ operator is highly conditioned. 
\par
The inverse of a mixing kernel is given with the expansion of the Neumann series
\begin{equation}
\label{eq:neuman}
(D + O)^{-1} = D^{-1} \sum_{k=0}^{\infty} (-D^{-1}O)^k = D^{-1} - D^{-1}OD^{-1} + D^{-1}OD^{-1}OD^{-1} - ...
\end{equation}
where the mixing was represented as a sum of its diagonal and off-diagonal elements.
%(the offdiag comes from coobserved mixing - hope this is clear for readers?). 
\par
The Chebyshev Graph Filter \cite{chebnet}
\begin{equation}
\label{eq:chebyshev}
y(L) = \sum_{m=0}^{M} c_m \, T_m(\tilde{L}) , 
\end{equation}
where $T_m$ are Chebyshev polynomials of the rescaled graph Laplacian $\tilde{L}$, and $c_m$ are learnable coefficients, corresponds to the summation in Equation \ref{eq:neuman}. The order $M$ in practice represents how many jumps over coobserved edges we want to consider in unmixing, and we limit it to three, since pixels further away cannot geometrically contribute to the same TOD sample. Furthermore, diagonal elements in Equation \ref{eq:neuman} are known. The mixing inversion is then embedded with 
\begin{equation}
\label{eq:inverseP}
\left( P^T P \right)^{-1} \approx \texttt{gnn.ChebFilter}(M=3)\, D^{-1}
= (P.T \mathbf{1})^{-1} \sum_{m=0}^{3} c_m \, T_m(\tilde{L})\, 
\end{equation}
where the Laplacian is built on coobserved edges. The inversion is efficient and interpretable, the coefficients represent how different secondary beams contribute to the reconstruction, the limitation $M$ represents the geometric limitation of the projection, and we implement the fact that the Neumann series is the closest approximation of the analytical solution. 

\section{Application to CMB Map Reconstruction}
All operators from Equation \ref{eq:inverseH} are implemented either deterministically (where possible) or approximately, as with examples from subsection \ref{subsec:examples}, directly into the architecture of a Sequential Neural Network \cite{sequential}. Different operators are kept as separate modules to allow individual work, which we here call \textit{modularity}. The proposed inversion network is applied to QUBIC TODs simulated with known sky maps. 
\par
\begin{figure}
    \centering
    \subfigure[]{\includegraphics[width=0.67\textwidth]{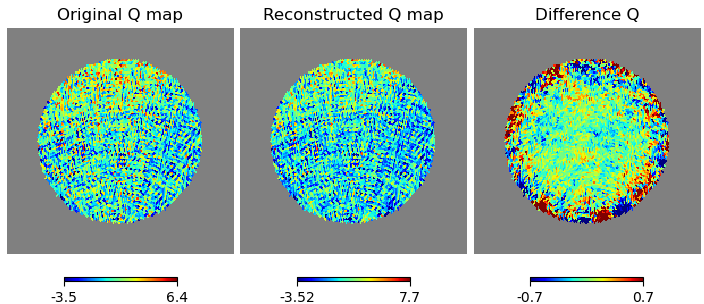}} 
    \subfigure[]{\includegraphics[width=0.67\textwidth]{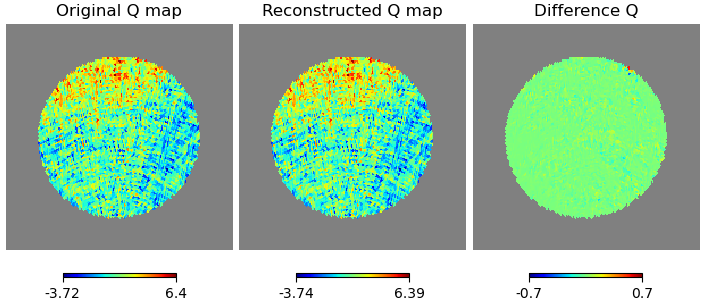}} 
    \caption{Simulated sky in Q Stokes, the corresponding reconstructed sky, and the reconstruction error. (a) Example reconstruction with the PCG method. (b) Example reconstruction with the proposed method. }
    \label{fig:rec}
\end{figure}

The operator $H^{T} H$ built from the full forward model in Equation \ref{eq:H} has the condition number of 43.3 for only one pointing at one frequency, while realistic scannings will span hundreds of thousands of pointings per hour. Separately treating the half-wave plate rotation with polarizer, bolometric decay, and the projection, reduces the condition number of the remaining operators to 1, indicating the solution is stable and well-defined.  
We compare our reconstruction with that of a standard PCG method. Our method achieves lower reconstruction error across all Stokes, particularly near the edge of the observed sky, shown in Figure \ref{fig:rec}. This improvement is due to the correct unmixing of coobserved pixels and the modular, individual approach to inverting the nondiagonal effects.

\par
In its simpler form, when layers are not used dynamically to estimate the instrument configuration constants, the network for CMB map reconstruction in multiple frequencies has fewer than 40 parameters. In the fitting mode, the number of parameters depends on the quantity of estimated constants, still being in the order of $\mathcal{O}(10)$. For comparison, a conventional ResNet architecture has around 11 million trainable parameters \cite{resnet}, and a classic U-Net for image segmentation typically contains about 30 million \cite{unet}.

\section{Conclusion}
In this work, we propose a modular, physics-guided approach in an effort to approximate the analytical solution of a severely ill-conditioned problem. Embedding the analytical inverses into the architecture, either deterministically or approximately, improves the quality of reconstruction while keeping the process interpretable, and allows for refining individual modules without the need for global retraining.  
\par
The results demonstrate that structured inversions, supported by the computational efficiency of the readily available deep learning packages, can improve the reconstruction in complex and ill-conditioned inverse problems. This approach offers a promising direction for building interpretable and physically consistent architectures in many fields as part of the emerging interest in the field of interpretable and explainable artificial intelligence.

%\section*{Acknowledgements}
%Acknowledgements should follow immediately after the conclusion.

% TODO: include author contributions
%\paragraph{Author contributions}
%This is optional. If desired, contributions should be succinctly described in a single short paragraph, using author initials.

% TODO: include funding information
%\paragraph{Funding information}
%Authors are required to provide funding information, including relevant agencies and grant numbers with linked author's initials. Correctly-provided data will be linked to funders listed in the \href{https://www.crossref.org/services/funder-registry/}{\sf Fundref registry}.

\bibliography{SciPost_Example_BiBTeX_File.bib}

%%%%%%%%%% END TODO: BIBLIOGRAPHY

\end{document}